\documentclass[manuscript]{aastex}
\usepackage{multirow}
\usepackage{color}

\newcommand{\msun}{\,M_\odot}
\newcommand{\kms}{\rm \,km\,s^{-1}}

\slugcomment{}

\shorttitle{Cloud-cloud collision which triggered formation of RCW38}
\shortauthors{Y. Fukui et al.}

\begin{document}

\title{The two molecular clouds in RCW\,38; evidence for formation of the youngest super star cluster in the Milky Way triggered by cloud-cloud collision}

\author{Y. Fukui\altaffilmark{1,2}, K. Torii\altaffilmark{1}, A. Ohama\altaffilmark{1}, K. Hasegawa\altaffilmark{1}, Y. Hattori\altaffilmark{1}, H. Sano\altaffilmark{1,2}, S. Ohashi\altaffilmark{3,4}, K. Fujii\altaffilmark{3,4}, S. Kuwahara\altaffilmark{3,4}, N. Mizuno\altaffilmark{4}, J. R. Dawson\altaffilmark{5,6}, H. Yamamoto\altaffilmark{1}, K. Tachihara\altaffilmark{1}, T. Okuda\altaffilmark{4}, T. Onishi\altaffilmark{7}, and A. Mizuno\altaffilmark{8}}
\affil{$^1$Department of Physics, Nagoya University, Chikusa-ku, Nagoya, Aichi 464-8601, Japan}
\affil{$^2$Institute for Advanced Research, Nagoya University, Chikusa-ku, Nagoya, Aichi 464-8601, Japan}
\affil{$^3$Department of Astronomy, School of Science, The University of Tokyo, 7-3-1 Hongo, Bunkyo-ku, Tokyo 133-0033, Japan}
\affil{$^4$National Astronomical Observatory of Japan, Mitaka, Tokyo 181-8588, Japan}
\affil{$^5$Department of Physics and Astronomy and MQ Research Centre in Astronomy, Astrophysics and Astrophotonics,
Macquarie University, NSW 2109, Australia}
\affil{$^6$Australia Telescope National Facility, CSIRO Astronomy and Space Science, P.O. Box 76, Epping, NSW 1710, Australia}
\affil{$^7$Department of Astrophysics, Graduate School of Science, Osaka Prefecture University, 1-1 Gakuen-cho, Nakaku, Sakai, Osaka 599-8531, Japan}
\affil{$^8$Solar-Terrestrial Environment Laboratory, Nagoya University, Chikusa-ku, Nagoya 464-8601, Japan}

\email{torii@a.phys.nagoya-u.ac.jp}

\begin{abstract}
We present distributions of two molecular clouds having velocities of 2$\kms$ and 14$\kms$ toward RCW\,38, the youngest super star cluster in the Milky Way, in the $^{12}$CO $J=$1--0 and 3--2 and $^{13}$CO $J=$1--0 transitions.
The two clouds are likely physically associated with the cluster as verified by the high intensity ratio of the $J$=3--2 emission to the $J$=1--0 emission, the bridging feature connecting the two clouds in velocity and their morphological correspondence with the infrared dust emission.
The velocity difference is too large for the clouds to be gravitationally bound. 
We frame a hypothesis that the two clouds are colliding with each other by chance to trigger formation of the $\sim$20 O stars which are localized within $\sim$0.5\,pc of the cluster center in the 2$\kms$ cloud. 
We suggest that the collision is currently continuing toward part of the 2$\kms$ cloud where the bridging feature is localized. 
This is the third super star cluster alongside of Westerlund\,2 and NGC\,3603 where cloud-cloud collision triggered the cluster formation. RCW\,38 is the youngest super star cluster in the Milky way, holding a possible sign of on-going O star formation, and is a promising site where we may be able to witness the moment of O-star formation. 

\end{abstract}


\section{Introduction}
\subsection{Important role of high-mass stars}
High-mass stars are very energetic and influential on dynamics of the interstellar medium via stellar winds, ultraviolet radiation and supernova explosions. Injection of the heavy elements at the end of their lives is another important effect of high-mass stars in galactic evolution. It is therefore one of the most important issues in astrophysics to understand formation of high-mass stars and considerable efforts have been made by a number of works published so far \citep[for recent reviews see][]{zin2007, tan2014}. 

\subsection{Theoretical attempts for understanding high-mass star formation}
It is now generally accepted that most of the high-mass stars are formed by mass accretion instead of collisional stellar merging after some years of debates. In the current paradigm, monolithic collapse \citep{yor2002} and competitive accretion \citep{bon2004} are the two scenarios of the mass accretion processes. In the monolithic collapse a massive dense molecular cloud is the initial condition for the star formation, which is followed by disk accretion, whereas in the competitive accretion a massive cluster of stars/pre-stellar cores is assumed to grow in mass by accretion of the ambient gas. The two scenarios still lack detailed confrontation with the initial conditions provided by observations. This is in part due to the lack of well-resolved observational views of high-mass star formation at small distances; even the Orion Nebular Cluster, the most outstanding high-mass star forming region at the smallest distance 400\,pc, is three times more distant than typical low-mass star-forming regions like Taurus, and the second nearests are at a distance around 2\,kpc. Therefore, our understanding of high-mass star formation yet remains to be significantly corroborated. 

It is interesting to note that insight-full discussion is made on the mechanisms of high-mass star formation in the review article by \citet{zin2007}, where we find the following sentence:
``Rapid external shock compression (i.e., supersonic gas motions) generating high-column densities in less than a local free-fall time rather than slow quasi-static build-up of massive cores may be the recipe to set up the initial conditions for local and global bursts of massive star formation.''
This idea of supersonic gas motions to collect gas was not explored further in the literature in the context of high-mass star formation except for the several papers on cloud-cloud collision discussed in Section 1.3.

\subsection{Increasing evidence for cloud-cloud collision as a triggering of high-mass star formation}
Molecular observations with NANTEN2 have shown that O-type star formation in the Milky Way disk is triggered by supersonic collisions between two clouds at velocity of 10\,--\,30$\kms$ in several places. The first discovery was made in the super star cluster Westerlund\,2 which is ionizing an H{\sc ii} region RCW\,49 \citep{fur2009,oha2010}, and the authors suggested that two giant molecular clouds collided to trigger formation of the cluster some 2\,Myrs ago. These observations were followed by discovery of cloud-cloud collisions which triggered formation of the super star cluster/mini-starburst in NGC\,3603 \citep{fuk2014}, and formation of the single O star in H{\sc ii} regions M\,20 \citep{tor2011} and RCW\,120 \citep{tor2015}. 
Most recently, in the Large Magellanic Cloud ALMA observations led to a discovery of formation of a 40\,$M_\odot$ star N159W-S triggered by collision between two filamentary clouds at a relative velocity of $\sim10\kms$ \citep[][]{fuk2015b}. These results raise a novel possibility that formation of O stars may be due to, or at least facilitated, by the supersonic compression in cloud-cloud collision, which was not considered in the theories for high-mass star formation, either in the monolithic collapse or in the competitive accretion scenarios. On a global scale of galaxies, numerical simulations show that collisions between clouds are fairly frequent at every $\sim$10\,Myrs \citep{tas2009, dob2015}, which is similar to an evolutionary timescale of giant molecular clouds \citep{fuk1999, kaw2009, fuk2010}.
It is thus becoming possible to consider cloud-cloud collision as one of the modes of high-mass star formation. 

Cloud-cloud collision was studied using hydrodynamics simulations by \citet{hab1992},  \citet{ana2010}, and \citet{tak2014}.
These authors found that cloud-cloud collision induces the formation of cloud cores by enhanced self-gravity as a consequence of shock compression, while the connection of cloud-cloud collision to high-mass star formation was not discussed into depth. Recent magnetohydrodynamical (MHD) numerical simulations on cloud-cloud collision show that colliding molecular gas can indeed create dense and massive cloud cores, precursors of high-mass stars, in the shock-compressed interface, giving a theoretical basis for triggered the O-star formation \citep{ino2013}. In the simulations after the collision the density increases from 300\,--\,1000\,cm$^{-3}$ to $\gtrsim10^4$\,cm$^{-3}$ and the turbulent velocity and magnetic field increase by about a factor of five, leading to a 100-times higher mass-accretion rate to form stars. 
This indicates that the conditions for the high mass-accretion rate postulated in high-mass star formation theories \citep{mck2003} are created by cloud-cloud collision.

Following these findings, as a next step we need to have a better understanding on the physical details of the collision, which include direct observations of the shock interaction as well as the statistical properties of cloud-cloud collisions.
We are carrying out a systematic study of high-mass star formation toward H{\sc ii} regions in order to elucidate the role of cloud-cloud collision with NANTEN2, Mopra, and ASTE telescopes. These H{\sc ii} regions include Spitzer bubbles, compact H{\sc ii} regions, and mini-starbursts including NGC\,6334, NGC\,6357, W\,43, etc. 

\subsection{RCW\,38 as the youngest super star cluster in the Milky Way}
Hoping to better understand the origin of massive clusters we in this paper focus ourselves on the super star cluster RCW\,38 \citep{rod1960}, which is the youngest super star cluster with an age of $\lesssim$1\,Myr in the Milky Way (\citealt{wol2006,wol2008}; see also Table\,2 in \citealt{por2010}). 
\\citet{wol2006} estimated the age of RCW\,38 as 0.5\,Myrs with an upper limit of 1\,Myrs so that it does not conflict with the number of O stars for the size of the cluster, and \citet{get2014} measured the ages of the several members of the RCW\,38 cluster, which ranges from 0.1\,--\,3.5\,Myrs.
The cluster age may be even smaller than 0.5\,Myrs as discussed in the present paper (Section 4).
The extreme youth will benefit study of the molecular gas close to the central O stars, which is particularly important to address the ionization/dispersal of the molecular gas. RCW\,38 has large total mass, similar to that of the Orion Nebula Cluster, and is an active high-mass star forming region, second nearest to the sun, at a distance of 1.7\,kpc. 
An overview of the previous studies on RCW\,38 is summarized in an article by \citet{wol2008}.

\begin{figure}
\epsscale{1.}
\plotone{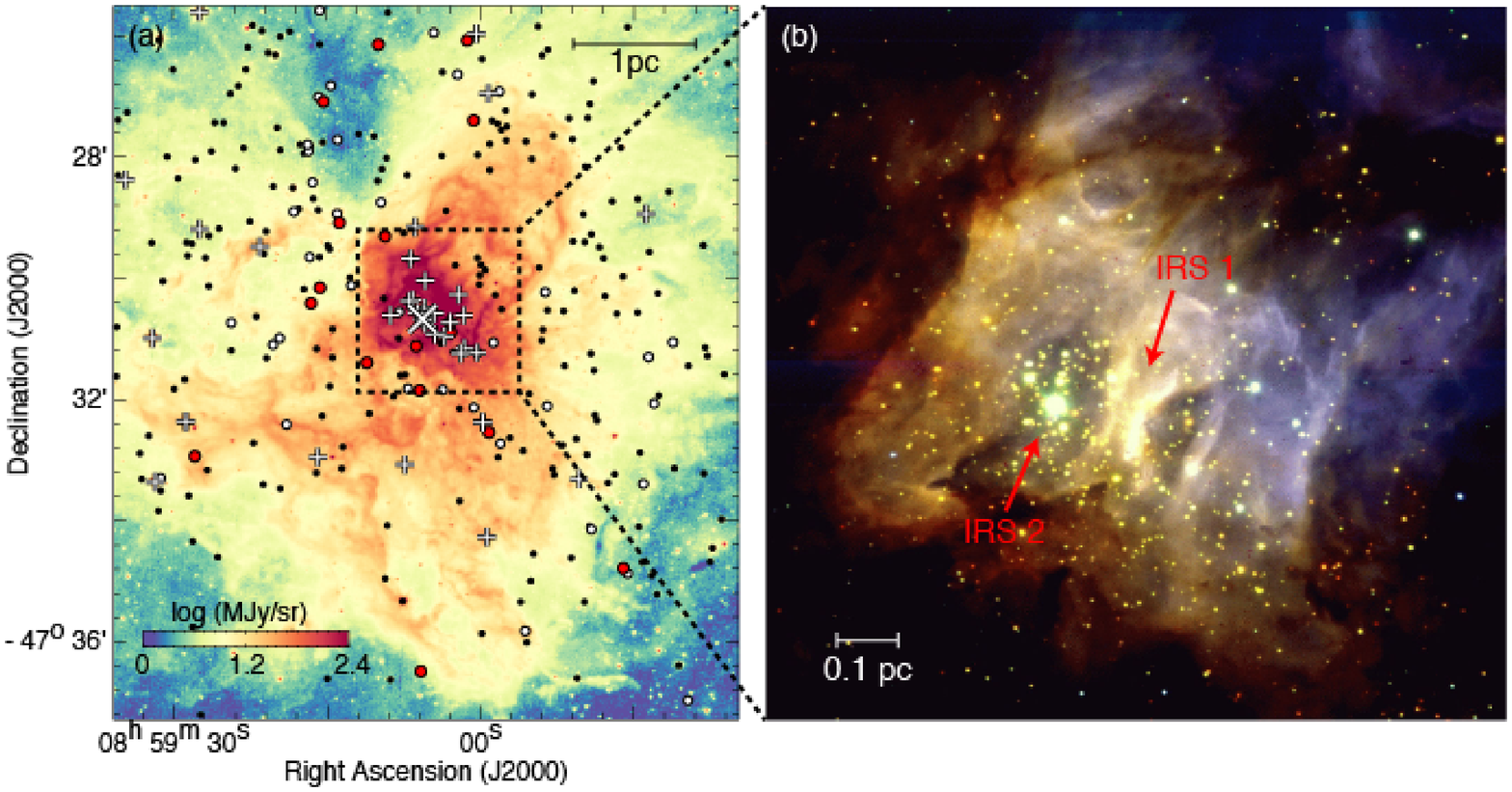}
\caption{(a) The 3.6\,$\mu$m image of RCW\,38 obtained with the Spitzer/IRAC observations \citep{wol2008}. YSOs and O star candidates obtained with \citet{wol2006} and \citet{win2011} are plotted in (b). Red circles and white circles indicate the class 0/I and flat spectrum YSOs, respectively, while black dots indicate the class II and III YSOs. White crosses indicate the candidate O stars.
(b) A close-up of the central region of RCW\,38 from the VLT observations \citep{wol2006}.  Z band is shown in blue, H band is green, and K band is red. The bright infrared emission IRS\,1 and IRS\,2 are shown by arrows.
 \label{rcw38}}
\end{figure}

RCW\,38 is known as one of the closest high-mass star forming regions containing $\sim$10,000 stars (\citealt{kuh2015, bro2013, wol2006}; see also \citealt{lad2003}) (Figure\,\ref{rcw38}). 
In the central part of RCW\,38, two remarkable infrared peaks have been identified \citep{fro1974}. 
The brightest at 2\,$\mu$m is labeled IRS\,2, which corresponds to an O5.5 binary located at the center of the RCW\,38 cluster \citep{der2009}. \citet{fur1975} derived a total infrared luminosity toward IRS\,2 to be $7\times10^5$\,$L_\odot$. 
The brightest feature at 10\,$\mu$m is found 0.1\,pc west of IRS\,2, and is labelled IRS\,1. 
IRS\,1 is a dust ridge extending by 0.1\,--\,0.2\,pc in the north-south direction (Figure\,\ref{rcw38}b), which has a color temperature of about 175\,K and includes several dust condensations \citep{smi1999}.
Infrared and millimeter wave observations has indicated a ``ring-like'' or ``horseshoe'' structure around IRS\,2 about 1$'$\,--\,2$'$ across \citep[][]{huc1974,vig2004,wol2006, wol2008}.
Inside the ring-like shape two regions are cleared of dust and form cavities; one is the region centered on IRS\,2 with a diameter of $\sim$0.1\,pc, and another is just west of IRS\,1 with a similar size \citep{smi1999, wol2008}, suggesting that they were formed by feedbacks from high-mass stars. 
A large-area infrared image shows that RCW\,38 includes numerous filaments and bubbles (see Figure\,\ref{rcw38}a). 
\citet{kan2013} made comparisons of the dust and polycyclic aromatic hydrocarbon (PAH) maps with the C[II] 158\,$\mu$m map and discussed that the PAHs and dust grains are well mixed along the line-of-sight and PAHs play an important role for photo-electric heating of gas in photo-dissociation regions.

Many observational studies to investigate the cluster members of RCW\,38 have been carried out so far.
Near infrared observations by using the Very Large Telescope (VLT) identified more than 300 young stars for the $\sim$0.5\,pc$^{2}$ area centered on IRS\,2 \citep{der2009}.
{\it Chandra} observations reported by \citet{wol2006} were used to identify 345 X-ray sources likely associated with RCW\,38, and completeness arguments suggest a total estimated cluster size of between 1500 and 2400 stars.
\citet{win2011} identified 624 young stellar objects (YSOs) by the {\it Spitzer}/IRAC observations toward a 30$'\,\times\,30'$ region of RCW\,38.
In a part of The Massive Young star-forming Complex Study in Infrared and X-rays (MYStIX), 886 pre-main sequence stars have been identified in RCW\,38, which are concentrated within 0.5\,pc of the cluster center \citep{fei2013, bro2013, kuh2013}.
About the high-mass stars, nearly 60 O-star candidates have been identified by \citet{wol2006} and \citet{win2011} in a large area of RCW\,38.
Out of them, about 20 O-star candidates are found in the central $\sim$0.5\,pc \citep{wol2006}.
\\citet{kuh2015} also pointed out that the RCW\,38 cluster shows a particularly high central stellar density compared with the other nearby clusters with O stars (e.g., the Orion Nebula Cluster, W40, RCW\,36, M17, etc.).
It seems natural that such an O star cluster has tens of O stars as cluster members, while only IRS\,2 is confirmed to be a double O5.5 star spectroscopically \citep{der2009}. We note cautiously that the number of O stars is still to be corroborated observationally by considering the uncertainty of the stellar mass determination based on pre-main-sequence models for the K-band stellar magnitude \citep{wol2006}.

Molecular line observations toward RCW\,38 are very limited.
\citet{yam1999} observed a large area of RCW\,38 in the $^{13}$CO $J$=1--0 emission with NANTEN and found a molecular cloud at 2\,--\,3$\kms$ which coincides with IRS\,2, although detailed structures including the ring-like component were not resolved due to a relatively coarse resolution. On the other hand, \citet{zin1995} observed the central region of RCW\,38 at HPBW of $\sim1'$ in the CS emission to resolve the ring shape surrounding IRS\,2.
$^{12}$CO $J$=1--0 observations with SEST at 45$''$ resolution for a small $160'' \times 160''$ region performed by \citet{gyu2008} show two molecular clouds likely associated with RCW\,38 at two different velocity ranges of $-3$\,--\,$+2\kms$ and $+3$\,--\,$+8\kms$.

In this paper, we present new molecular line observations toward RCW\,38 obtained with NANTEN2, Mopra, and ASTE telescopes. 
The NANTEN2 CO $J$=1--0 observations covered a large area of RCW\,38 at an angular resolution of 4$'$ to describe the large scale molecular morphologies and velocities, while the Mopra CO $J$=1--0 and ASTE  CO $J$=3--2 observations at angular resolutions of 22$''$\,--\,35$''$ revealed the detailed molecular gas distributions and physical conditions such as temperature and density around IRS\,1 and IRS\,2.
Section 2 describes observations and Section 3 observational results and the physical properties of the molecular clouds. Section 4 gives discussion on cloud-cloud collision and O star formation, and Section 5 concludes the paper.

\section{Observations}
\subsection{NANTEN2 CO $J$=1--0 observations}
The NANTEN2 4-m mm/sub-mm telescope situated in Chile was used to observe a large area $1.8^\circ \times 1.8^\circ$ of RCW\,38 in the $^{12}$CO and $^{13}$CO $J$=1--0 transitions during a period from May 2012 to December 2012.
The 4\,K cooled SIS mixer receiver equipped with NANTEN2 provided a typical system temperature $T_{\mathrm{sys}}$ of $\sim$250\,K in DSB.
The backend was a digital spectrometer which provided 16384 channels at 1\,GHz bandwidth and 61\,kHz resolution, corresponding to 2600$\kms$ and 0.17$\kms$, respectively, at 110\,GHz. 
The obtained data were smoothed to a velocity resolution of 0.8$\kms$ and angular resolution of 240$''$.
The pointing accuracy was checked to be better than 15$''$ with daily observations toward the Sun and IRC\,10216 (R.A., Dec.)=($9^{\rm h}\,47^{\rm m}\,57\fs406$, $13\degr16\,\arcmin\,43\farcs56$). The equinox of the celestial coordinates used in this paper is J2000.0.
The absolute intensity calibration was done with daily observations of the $\rho$-Ophiuchus molecular cloud (R.A., Dec.)=($16^{\rm h}\,32^{\rm m}\,23\fs3$, $-24\degr\,28\arcmin\,39\farcs2$) and Perseus molecular cloud (R.A., Dec.)=($3^{\rm h}\,29^{\rm m}\,19\fs0$, $31\degr24\,\arcmin\,49\farcs0$). 

\subsection{Mopra CO $J$=1--0 observations}
Detailed CO $J$=1--0 distributions around RCW\,38 were obtained by using the 22-m ATNF (Australia Telescope National Facility) Mopra mm telescope in Australia at angular resolution of 33$''$ in July 2012. 
We simultaneously observed the $^{12}$CO $J$=1--0, $^{13}$CO $J$=1--0, and C$^{18}$O $J$=1--0 transitions toward a $11' \times 15'$ area of RCW\,38 in the OTF mode with a unit field of $4'\times4'$.
$T_{\mathrm{sys}}$ was 400\,--\,600 K in the SSB. 
The Mopra backend system ``MOPS" which provided 4096 channels across 137.5 MHz in each of the two orthogonal polarizations was used in the observations. The velocity resolution was 0.088$\kms$ and the velocity coverage was 360$\kms$ at 115\,GHz. 
The obtained data were smoothed to a HPBW of 40$''$ with a 2D Gaussian function and to 0.6$\kms$ velocity resolution. 
The pointing accuracy was checked every 1 hour to keep within 7$''$ with observations of 86\,GHz SiO masers.
We made daily observations of Orion-KL (R.A., Dec.)=($-5^{\rm h}35^{\rm m}14\fs5$, $-5\degr22\arcmin29\farcs6$) to estimate ``extended beam efficiency" by making comparisons with the peak temperature of 100\,K in Orion-KL shown in Figure 6 of \citet{lad2005}. 
We finally obtained an extended beam efficiency of 0.48.

\subsection{ASTE $^{12}$CO $J$=3--2 observations}
Observations of RCW\,38 in the $^{12}$CO $J$=3--2 transition were performed by using the ASTE 10-m telescope located in Chile in 2013 \citep{eza2004, eza2008}. The waveguide-type sideband-separating SIS mixer receiver for the single sideband (SSB) ``CAT345'' having system temperature of $\sim250$\,K \citep{ino2008} and the digital spectrometer ``MAC'' with the narrow-band mode providing 128\,MHz bandwidth and 0.125\,MHz resolution \citep{sor2000}, which correspond to 111$\kms$ velocity coverage and 0.11$\kms$ velocity resolution at 345\,GHz, were used. 
The observations were made with the OTF mode at a grid spacing of 7.5$''$, and the HPBW was 22$''$ at the $^{12}$CO $J$=3--2 frequency.
The pointing accuracy was checked every $\sim1.5$\,hours and kept within 5$''$ with observations of RAFGL\,5254 (R.A., Dec.)=($09^{\rm h}13^{\rm m}53\fs94$, $-24\degr51\arcmin25\farcs1$).
The absolute intensity calibration was done with observations of Orion-KL (R.A., Dec.)=($-5^{\rm h}35^{\rm m}14\fs5$, $-5\degr22\arcmin29\farcs6$), and the day-to-day fluctuations of the peak intensity were within 10\,\%.

\section{Results}\label{sec:results}
\subsection{Large-scale molecular distributions with NANTEN2}
We first present large-scale molecular distributions of RCW\,38 using the NANTEN2 $^{12}$CO $J$=1--0 dataset.
Compared with the NANTEN observations given by \citet{yam1999}, although the beam size of 180{\arcsec} is the same, a grid spacing of 60{\arcsec} in the present NANTEN2 observations is a third of the size, which enables us to reveal the gas distributions into more detail.

\begin{figure}
\epsscale{1.}
\plotone{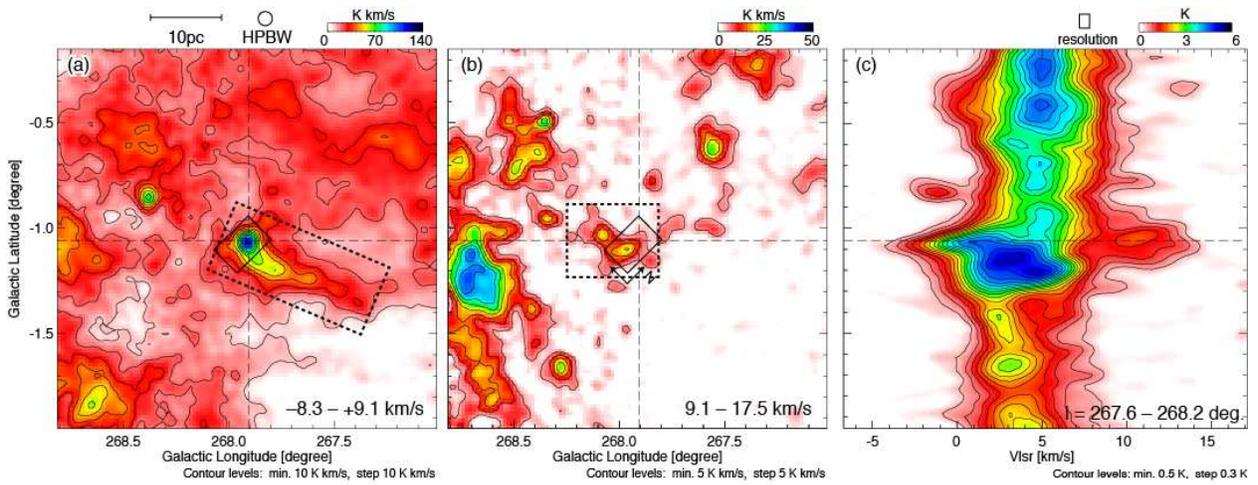}
\caption{$^{12}$CO $J$=1--0 integrated intensity distributions of the ring cloud (a) and the finger cloud (b). The area presented in Figure\,\ref{lb+vb} with the Mopra data is shown by black lines in (b). (c) Latitude-velocity diagram of the $^{12}$CO $J$=1--0 emission. Dashed line indicates the position of RCW\,38 IRS\,2, while thick dashed lines show the regions used for the molecular mass estimates.
 \label{nasco_lb}}
\end{figure}

As shown in the CO distributions in Figure\,\ref{nasco_lb}, toward the direction of RCW\,38, two CO clouds are distributed at radial velocities $v_{\rm LSR}$ of $-4$\,--\,+8$\kms$ and +9\,--\,+14$\kms$.
In Figure\,\ref{nasco_lb}(a), the blue-shifted cloud has a strong peak just coinciding with the central part of RCW\,38.
The compact peak with a size of 0{\fdg}1 ($\sim3$\,pc at RCW\,38) has a feature elongated toward the southwest of RCW\,38, showing a ``head-tail'' structure, which was also detected in \citet{yam1999}.
The head-tail structure is surrounded by the diffuse CO emission distributed at around $v_{\rm LSR}$ of $0$\,--\,+8$\kms$, covering the present observed region.

Contrary to the blue-shifted cloud shown in Figure\,\ref{nasco_lb}(a), the red-shifted cloud has weak and dispersed CO distribution. 
A CO peak at $(l, b)$\,$\simeq$\,$(268{\fdg}0, -1{\fdg}1)$ with relatively strong emission is located just at the southeast of RCW\,38, pointing toward the center of RCW\,38. 
This CO peak is well covered with the Mopra and ASTE observations as indicated by a box with solid lines in Figure\,\ref{nasco_lb}(b).
There are several other CO components around the CO peak, and they all seem to be continuously distributed along the east-west direction roughly at a range of $(l,b)\sim(267{\fdg}5\,--\,268{\fdg}4, -1{\fdg}0)$. 
In the Dec.-velocity diagram in Figure\,\ref{nasco_lb}(c), the red-shifted cloud can be identified separately from, although it looks being connected with, the blue-shifted cloud and the diffuse emission, suggesting a possible physical relationship of the red-shifted cloud with RCW\,38.

\subsection{Detailed molecular distributions with Mopra and ASTE}
\begin{figure}
\epsscale{1.}
\plotone{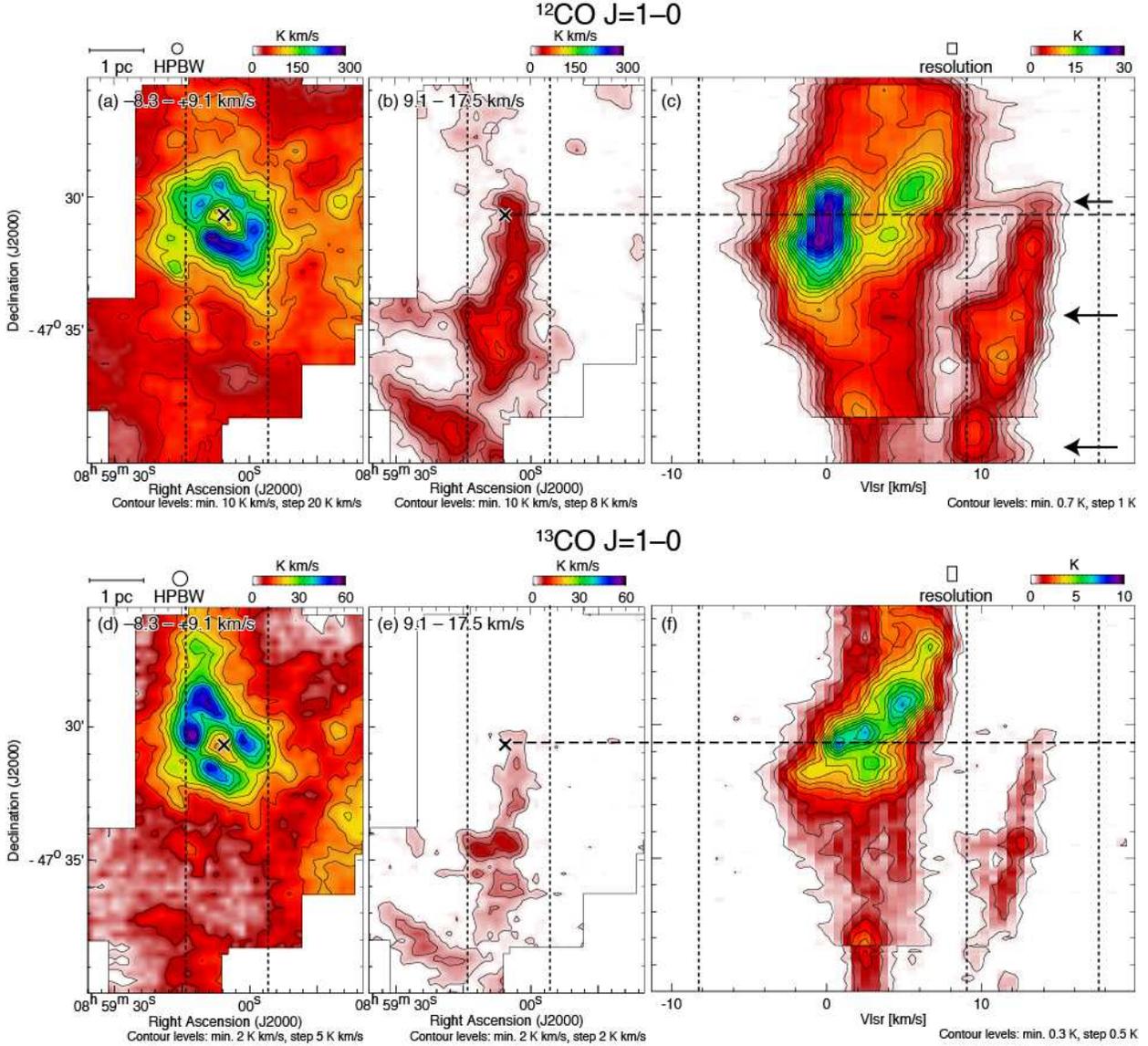}
\caption{Integrated intensity distributions of the ring cloud and the finger cloud are shown in the left and central four panels, and the Dec.-velocity diagrams are in the right two panels. Upper three panels show the $^{12}$CO $J$=1--0 emission, and lower three shows the $^{13}$CO $J$=1--0 emission. In (a), (b), (d), and (e), crosses show the position of IRS\,2, while the vertical dashed
lines indicate the integration range in the declination-velocity diagram in (c) and (f). Thin dashed lines indicate the observed area with Mopra. In (c) and (f) the vertical dashed lines indicate the integration ranges in the integrated intensity maps in (a), (b), (d), and (e). Thick horizontal dashed lines indicate the declination of IRS\,2. Three arrows in (c) indicate the declinations of the bridge features, BR1, BR2, and BR3, shown in Figure\,\ref{comp}.
 \label{lb+vb}}
\end{figure}

\begin{figure}
\epsscale{.4}
\plotone{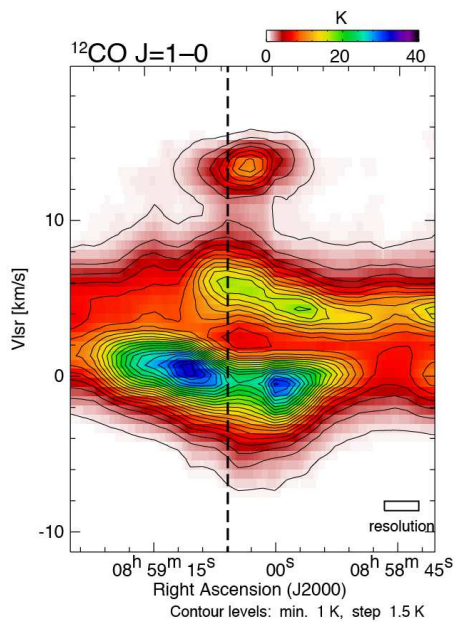}
\caption{R.A.-velocity diagram of $^{12}$CO $J$=1--0 is shown. Integration range in Dec. is from $-47\degr\,32\farcm2$ to $-47\degr\,29\farcm4$. Vertical dashed line indicates the direction of IRS\,2.
 \label{lv}}
\end{figure}

\begin{figure}
\epsscale{.75}
\plotone{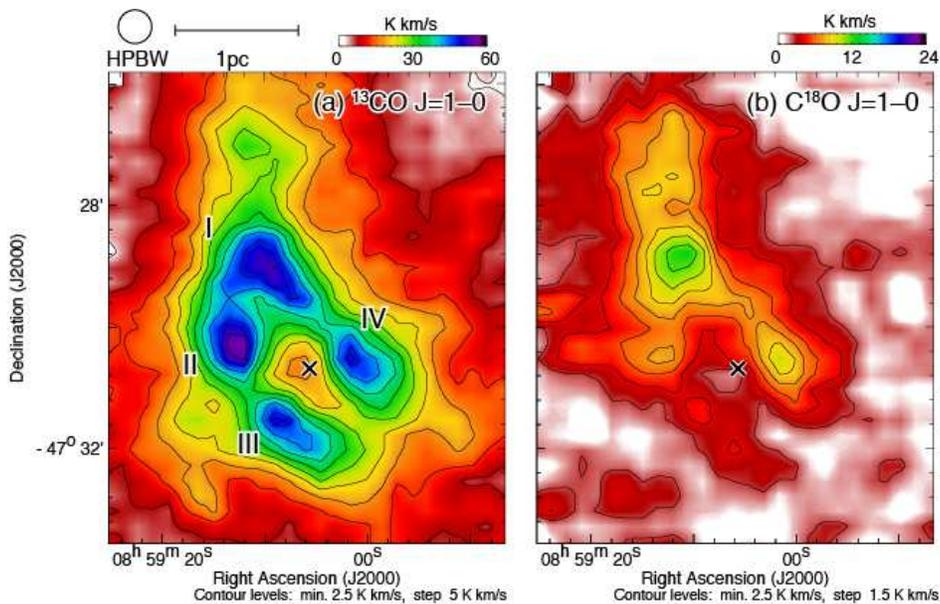}
\caption{Enlarged integrated intensity distributions of the ring cloud in $^{13}$CO $J$=1--0 (a) and C$^{18}$O $J$=1--0 for a velocity range of $-8.3$\,--\,$+9.1$$\kms$. Crosses indicate the position of IRS\,2.
 \label{mopra13+18}}
\end{figure}

We present the detailed molecular distributions around RCW\,38 using the Mopra and ASTE datasets.
In Figure\,\ref{lb+vb}, the integrated intensity distributions (left and central panels) and the Dec.-velocity diagram (right panels) are shown for the $^{12}$CO $J$=1--0 emission (upper panels) and the $^{13}$CO $J$=1--0 emission (lower panels).
The R.A.-velocity diagram in the $^{12}$CO $J$=1--0 emission is also shown in Figure\,\ref{lv}.
In Figures\,\ref{lb+vb}(a) and \ref{lb+vb}(d), the outstanding compact peak seen in Figure\,\ref{nasco_lb}(a) is resolved.
It shows a ring-like structure having a diameter of 1\,--\,2\,pc, with a cavity centered on IRS\,2, and several filamentary and bubble-like structures are radially extended outside the ring-like structure by 1\,--\,2\,pc. 
In the $^{12}$CO position-velocity diagrams in Figures\,\ref{lb+vb}(c) and \ref{lv}, intensity decrease is seen at around $v_{\rm LSR}$ of 2\,--\,4$\kms$. 
The $^{13}$CO emission in Figure\,\ref{lb+vb}(f) shows a single peak at $\sim2\kms$, indicating that the dip at the same velocity in the $^{12}$CO is due to self-absorption\footnote{The two velocity features reported by \citet{gyu2008} in $^{12}$CO $J$=1--0 correspond to the two peaks caused by self-absorption.}
The blue-shifted cloud has a peak at 2$\kms$.
The ring-like structure in $^{13}$CO consists of four clumps (clumps I\,--\,IV), and the ring-like structure itself has an elongated shape along the northeast-southwest direction (Figure \ref{mopra13+18}). 
The size of the clumps ranges from 0.3\,pc to 0.5\,pc, and the linewidth $\Delta v$ from 4$\kms$ to 5$\kms$ (Table \ref{table_clump}).

\begin{figure}
\epsscale{.45}
\plotone{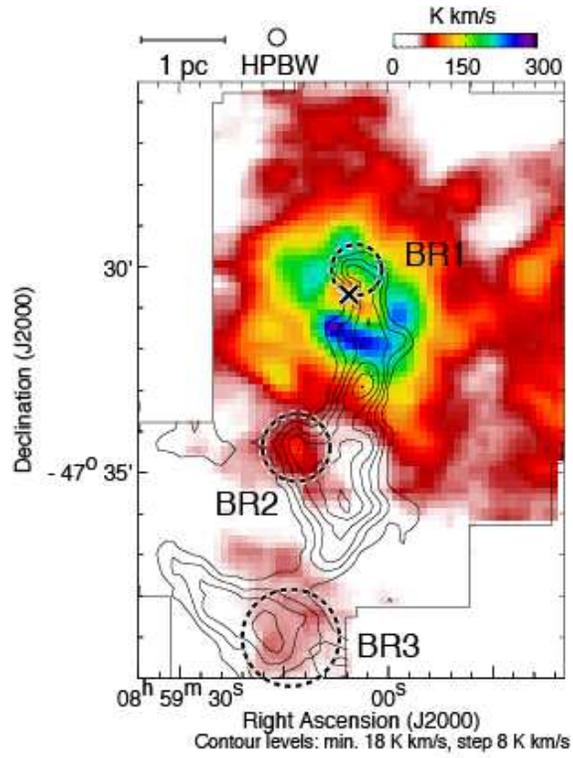}
\caption{A comparison of the $^{12}$CO $J$=1--0 distributions between the ring cloud (color) and the finger cloud (contours) is presented. BR1, BR2, and BR3 show the positions of the bridging features by dashed circles.
 \label{comp}}
\end{figure}

In contrast to the complicated distributions of the 2$\kms$ cloud, the 14$\kms$ cloud has a simple ridge elongated from the north to the south (Figures\,\ref{lb+vb}b and \ref{lb+vb}e).
We shall hereafter call the 2$\kms$ cloud and the 14$\kms$ cloud as the {\it ring cloud} and the {\it finger cloud}, respectively, by their shapes.
The finger cloud is not likely affected by self-absorption in $^{12}$CO, since $^{12}$CO and $^{13}$CO show similar distributions in the Dec.-velocity maps in Figures\,\ref{lb+vb}(c) and \ref{lb+vb}(f).
We find three ``bridging'' features in velocity between the ring cloud and the finger cloud at Dec.\,$\simeq$\,$-47{\degr}\,30{\farcm}0$\,(toward the cluster), $-47{\degr}\,34\farcm5$ and $-47{\degr}\,39\farcm5$, respectively, as indicated by arrows in Figure \ref{lb+vb}(c) and they are not seen beyond the velocity of the finger cloud.
Their positions are denoted as BR1, BR2 and BR3 in Figure \ref{comp}. The bridging feature distributed toward the cluster is also seen in the R.A.-velocity diagram in Figure\,\ref{lv}.
At its bottom, the ridge intersects another elongated structure running from the east to the west, and these two orthogonal elongated structures comprise the finger cloud in the present observed area (Figures\,\ref{lb+vb}(b) and (e)).
As shown in Figures\,\ref{lb+vb}(c) and \ref{lb+vb}(f), the finger cloud shows a uniform velocity gradient of $\sim$1$\kms$\,pc$^{-1}$ along the ridge.

\subsection{Intensity ratios of the $^{12}$CO $J$=3--2 emission to the $^{12}$CO $J$=1--0 emission} \label {sec:intensityratio}
In order to investigate the physical properties of the ring cloud and the finger cloud, we present distributions of the intensity ratio of the $^{12}$CO $J$=3--2 emission to the $^{12}$CO $J$=1--0 emission (hereafter $R_{3-2/1-0}$). 
Taking intensity ratios between different $J$-levels of CO is a useful 
diagnose of the molecular gas properties \citep[e.g.,][]{oha2010, tor2011, fuk2014}.

\begin{figure}
\epsscale{.7}
\plotone{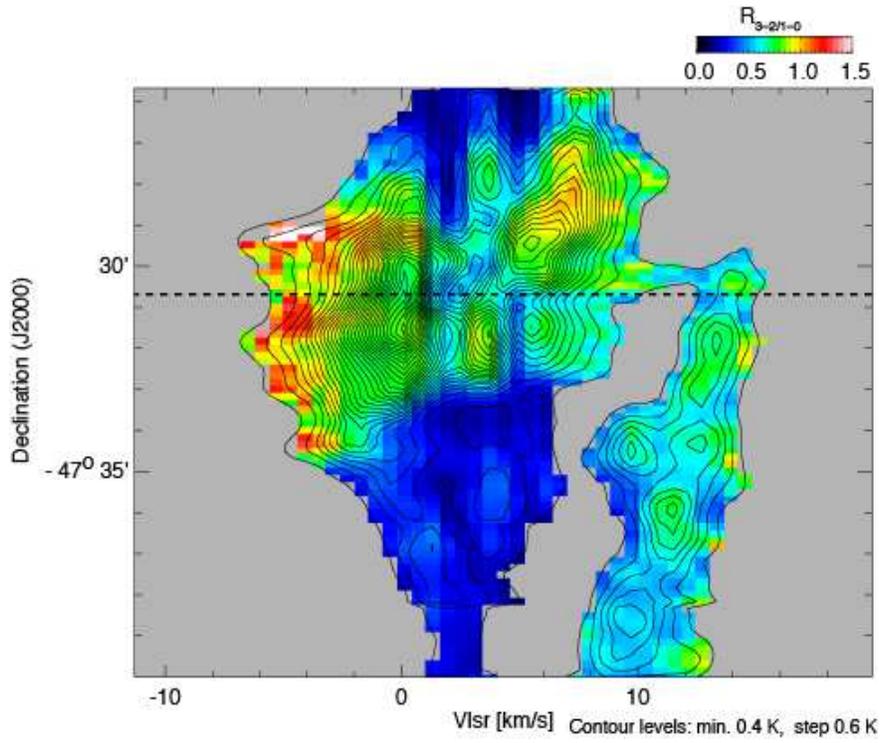}
\caption{The Dec.-velocity diagram of the $^{12}$CO $J$=3--2/$J$=1--0 ratio. The integration range in the R.A. is the same as Figure\,\ref{lb+vb}(c). The dashed line indicates the position of IRS\,2.
 \label{ratio_vb}}
\end{figure}

As seen in the Dec.-velocity diagram of $R_{3-2/1-0}$ in Figure\,\ref{ratio_vb}, both of the ring cloud and the finger cloud have typical ratios of 0.6\,--\,0.8, up to over 1.0. 
Since the ring cloud is strongly affected by the self-absorption in $^{12}$CO spectra around its central velocity range, $\sim$0\,--\,4$\kms$, the ratio in this velocity range is not reliable.
On the other hand, low intensity ratios are seen in the northern part (Dec.\,$>$\,$-47{\degr}\,27\farcm0$) and the southern part (Dec.\,$<$\,$-47{\degr}34{\farcm}0$) of the ring cloud, where the diffuse CO emission is widely distributed, while the finger cloud retains its relatively high intensity ratios of $\sim$0.6\,--\,0.8 throughout the cloud.

\subsection{Temperature and density of the molecular gas}
\begin{figure}
\epsscale{.6}
\plotone{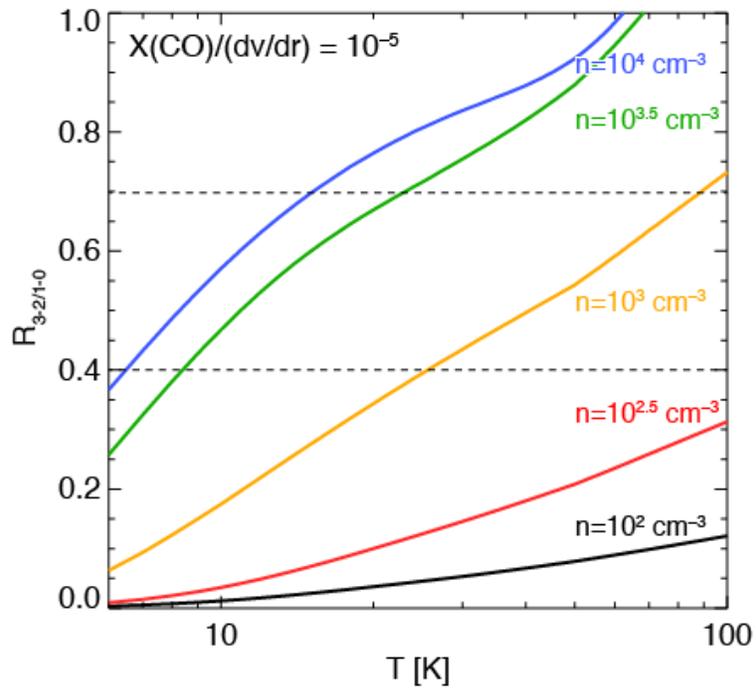}
\caption{Curves of the $^{12}$CO $J$=3--2/$J$=1--0 ratio as a function of $T_{\rm k}$ for various $n$(H$_2$) ranges estimated with LVG for $X({\rm CO})/(dv/dr)$ of $10^{-5}$\,(km\,s$^{-1}$\,pc$^{-1}$)$^{-1}$. 
 \label{lvg}}
\end{figure}

We utilize the large velocity gradient (LVG) analysis \citep[e.g.,][]{gol1974} to estimate kinetic temperature $T_{\rm k}$ and molecular number density $n$(H$_2$) of the two clouds in order to interpret their intensity ratio distributions.
We first present curves of $R_{3-2/1-0}$ as a function of $T_{\rm k}$ in Figure\,\ref{lvg} for various densities from $10^{2}$\,cm$^{-3}$ to $10^4$\,cm$^{-3}$, where $X({\rm CO})/(dv/dr) = 10^{-5}$\,(km\,s$^{-1}$\,pc$^{-1}$)$^{-1}$ is assumed.
We adopt an abundance ratio $X({\rm CO})$ = [$^{12}$CO]/[H$_2$] = $10^{-4}$ \citep[e.g.,][]{fre1982, leu1984}, and typical $dv/dr$ is estimated for clumps I\,--\,IV as 4$\kms$/0.4\,pc\,=\,10$\kms$\,pc$^{-1}$.

Figure\,\ref{lvg} provides a guide for the $R_{3-2/1-0}$ distributions in Figures\,\ref{ratio_vb}.
If $R_{3-2/1-0}$ is larger than 0.7, which is depicted by a dashed line in Figure\,\ref{lvg}, $T_{\rm k}$ is always higher than 10\,K for every $n$(H$_2$).
10\,K is a canonical value of the galactic molecular gas without significant star formation, suggesting that the gas with $T_{\rm k}$ higher than 10\,K is caused by some additional heating.
On the other hand, if $R_{3-2/1-0}$ is as small as 0.4, $T_{\rm k}$ cannot be determined, since it has a large variation from $<$10\,K to $>$100\,K depending on $n$(H$_2$).
Both of the ring cloud and the finger cloud show $R_{3-2/1-0}$ higher than 0.7 up to over 1.0 at many points, suggesting that these two clouds have $T_{\rm k}$ higher than 10\,K.
It is reasonable that the heating is mainly due to the O stars in RCW\,38.
On the other hand, the diffuse CO emission surrounding the ring cloud with $R_{3-2/1-0}$ of 0.4 is of low excitation.

\begin{figure}
\epsscale{.9}
\plotone{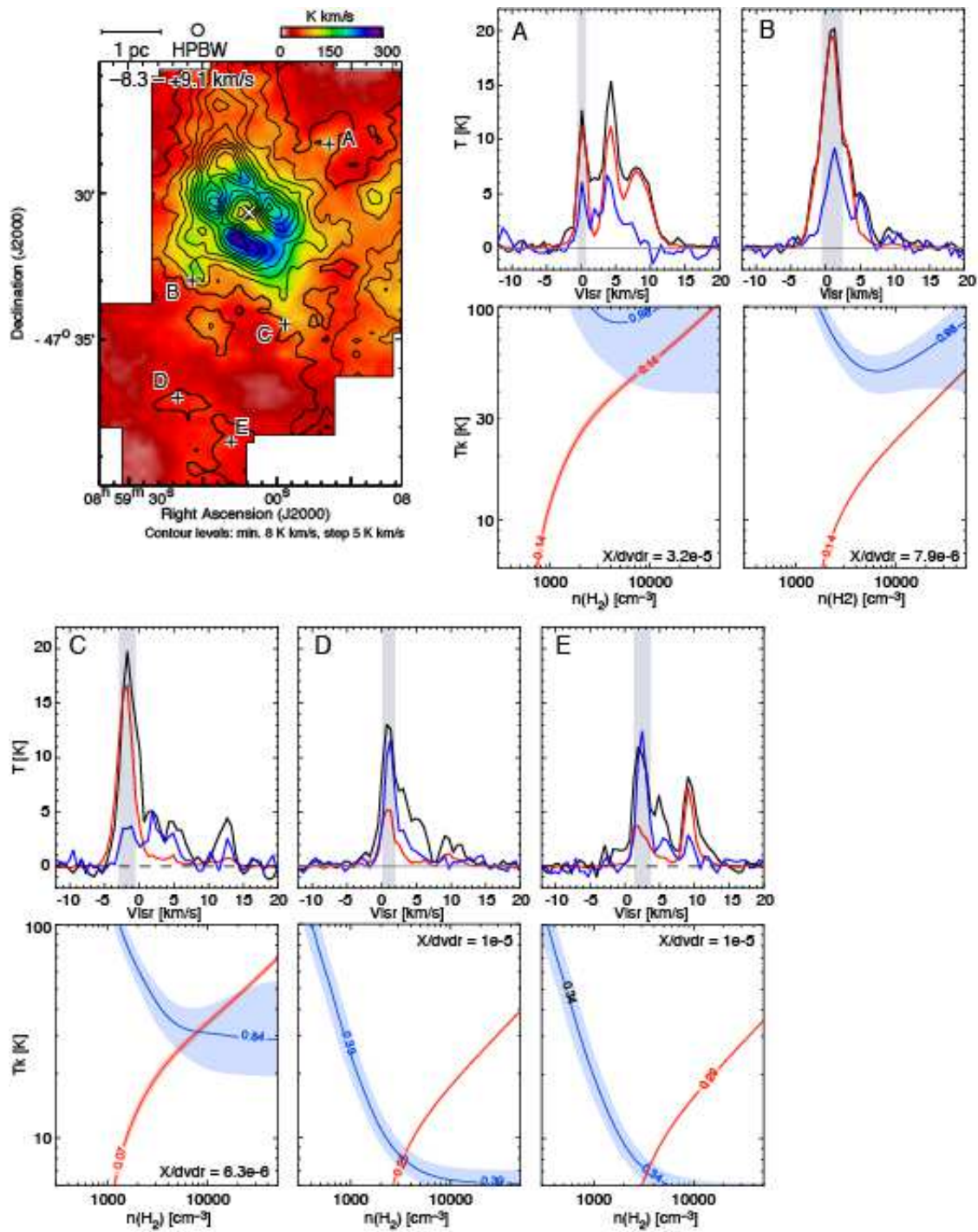}
\caption{LVG results on the $n$(H$_2$)-$T_{\rm k}$ plane for the regions A--E in the ring cloud are shown. $^{12}$CO $J$=3--2/$J$=1--0 is plotted in black, and $^{13}$CO/$^{12}$CO $J$=1--0 is in red. The regions A--E are plotted with black crosses in the CO integrated intensity map in the upper-left panel, where $^{12}$CO $J$=3--2 is shown in color, while $^{13}$CO $J$=1--0 is shown with contours. White cross indicates the position of IRS\,2. Spectra of $^{12}$CO $J$=1--0 (black), $^{12}$CO $J$=3--2 (red), and $^{13}$CO $J$=1--0 (blue) are also shown. Intensities of the $^{13}$CO $J$=1--0 spectra are multiplied by three. 
 \label{lvg1}}
\end{figure}

\begin{figure}
\epsscale{.9}
\plotone{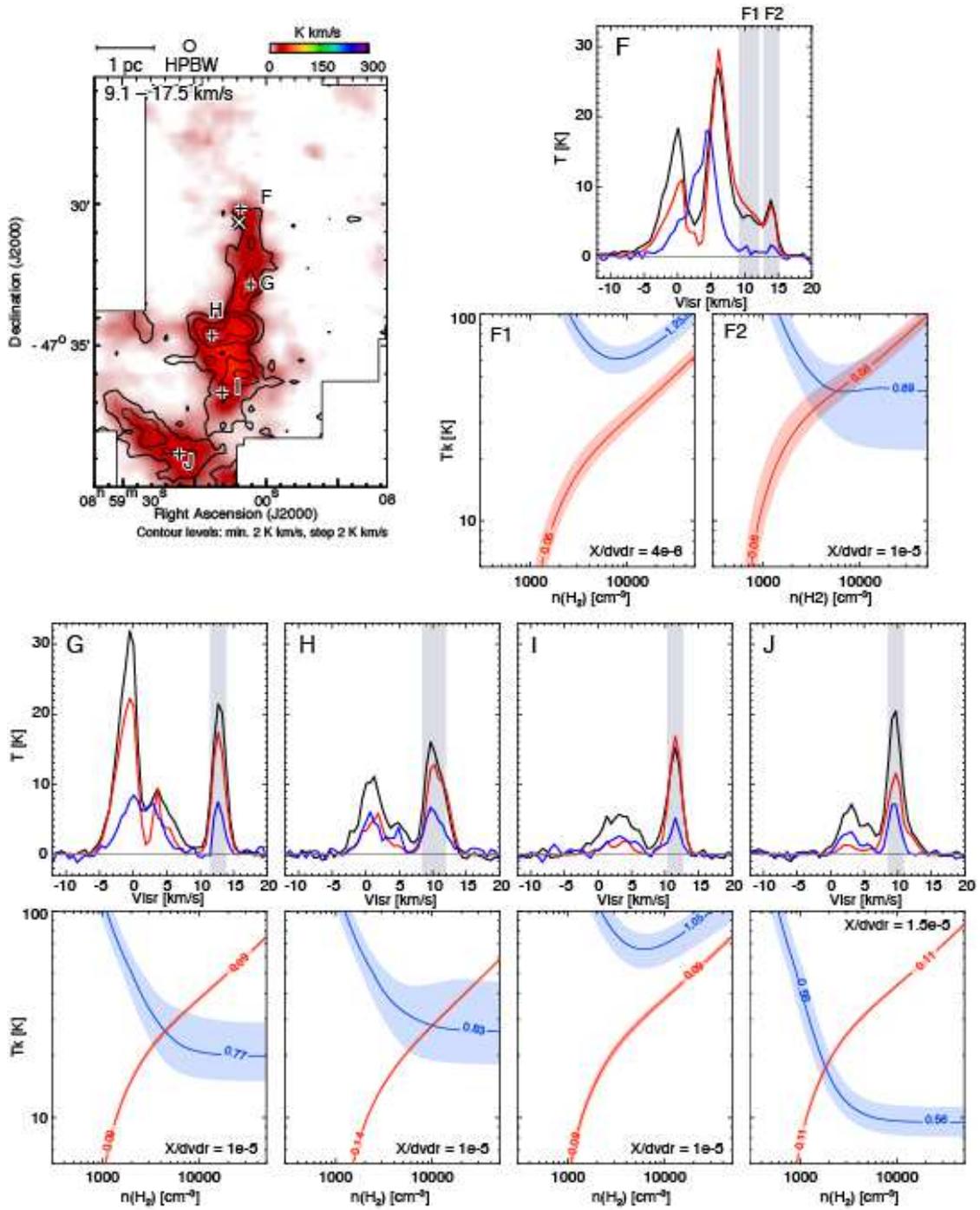}
\caption{LVG results for the regions F--J in the finger cloud in the same manner as Figure\,9.
 \label{lvg2}}
\end{figure}

In order to provide more quantitative constraints for $T_{\rm k}$ and $n$(H$_2$), we add the ratio of $^{13}$CO $J$=1--0 to $^{12}$CO $J$=1--0 (hereafter $R_{13/12}$) in the LVG analysis, with an assumption of the abundance ratio [$^{12}$CO]/[$^{13}$CO] = 77 \citep{wil1994}.
Ten target regions A\,--\,J for $T_{\rm k}$ and $n$(H$_2$) estimates in the two clouds.
The CO spectra and the LVG results are presented in Figures\,\ref{lvg1} and \ref{lvg2} for the individual target regions.
In the CO spectra, the velocity ranges used for the LVG analysis are shown by shade.
For the ring cloud, the five target regions are chosen to have apparent self-absorption in their $^{12}$CO spectra.
For the ring cloud, regions A\,--\,C are taken at the high $R_{3-2/1-0}$ regions around the rim of the dense part, while regions D and E are at the low $R_{3-2/1-0}$ regions widely distributed in the north and the south.
Regions F\,--\,J in Figure\,\ref{lvg2} are taken to cover a large area of the finger cloud.
In the diagrams of the LVG results, R$_{3-2/1-0}$ and $R_{13/12}$ distributions are shown with the blue lines and red lines, respectively, and their errors are shown with the colored area.
The errors are estimated with 1\,$\sigma$ noise fluctuations of the spectra and the 10\,\% relative calibration error for each CO transition. 
Since the $^{12}$CO $J$=1--0 emission and the $^{13}$CO $J$=1--0 emission were taken simultaneously with the same receiver and the same backend at Mopra, we assume that the 10\,\% relative calibration error is canceled for $R_{13/12}$ and adopt the error only for $R_{3-2/1-0}$.
In addition, $dv/dr$ is estimated for the individual regions, and $X({\rm CO})$ is assumed to be $10^{-4}$ same as in Figure\,\ref{lvg}.
Finally, $T_{\rm k}$ and $n$(H$_2$) are given at the region where R$_{3-2/1-0}$ curve and $R_{13/12}$ curve overlap.

The results are summarized as follows:
Regions A\,--\,C in the ring cloud and all the regions in the finger cloud (regions F\,--\,J) show $T_{\rm k}$ higher than 10\,K, typically 30\,--\,40\,K, up to more than 50\,K.
Typical $n$(H$_2$) for these regions are about $10^3$\,--$10^4$\,cm$^{-3}$, where only lower limits are given for regions A, B, and F2 and no solutions are given for regions F1 and I.
Regions D and E located in the south of the ring cloud show significantly low $T_{\rm k}$ less than 10\,K and $n$(H$_2$) of $\sim3\times10^3$\,cm$^{-3}$.

As a summary, both of the ring cloud and the finger cloud show remarkably high temperatures of $T_{\rm k} > 20$\,--\,30\,K.
Some heating source/mechanism is necessary to understand the results, and radiative heating by the O stars in RCW\,38 is a reasonable mechanisum.
The present results suggest that, despite of a large velocity separation of $\sim$12$\kms$, both of the ring cloud and the finger cloud are associated with RCW\,38.

\subsection{Molecular masses of the ring cloud and the finger cloud}
By adopting the distance of RCW\,38, 1.7\,kpc, to both the ring cloud and the finger cloud, their molecular masses are estimated.
For a large scale, the NANTEN2 $^{12}$CO $J$=1--0 data presented in Figures\,\ref{nasco_lb}(a) and \ref{nasco_lb}(b) are used to estimate the total molecular masses of the two clouds. 
We use an $X$(CO) factor of $2\times10^{20}$\,cm$^{-2}$\,(K$\kms$)$^{-1}$ \citep{str1988}, which is an empirical conversion factor from $^{12}$CO $J$=1--0 integrated intensity into H$_2$ column density. 
The masses estimated with the X(CO) factor (hereafter $M_{X({\rm CO})}$) of the head-tail structure of the ring cloud and the finger cloud are calculated as $3.0\times10^4$$\msun$ and $1.9\times10^3$$\msun$ toward the regions enclosed with dashed lines in Figure\,\ref{nasco_lb}(a) and \ref{nasco_lb}(b), respectively.
$M_{X({\rm CO})}$ of the finger cloud is more than one order of magnitude smaller than that of the ring cloud.

The molecular masses of the central region of RCW\,38 shown in Figure\,\ref{lb+vb}(b) are estimated with the Mopra CO data set.
$M_{X({\rm CO})}$ of the finger cloud is estimated to be $1.2\times10^3$$\msun$ at $^{12}$CO $J$=1--0 integrated intensity $W$($^{12}$CO\,1--0)\,$\geq$\,10\,K$\kms$ ($= 3\,\sigma$).
On the other hand, because the $^{12}$CO $J$=1--0 intensity is reduced by by the strong self-absorption particularly in the central 1\,--\,2\,pc of the ring cloud (see Figure\,\ref{lb+vb}(c)), we estimate the mass of the ring cloud using the Mopra $^{13}$CO $J$=1--0 data assuming the local thermodynamic equilibrium (LTE).
In estimating the molecular mass with the LTE assumption (hereafter $M_{\rm LTE}$), excitation temperature $T_{\rm ex}$ must be provided, and the peak intensity of the $^{12}$CO $J$=1--0 emission $T(^{12}{\rm CO})$ is usually used to estimate $T_{\rm ex}$ with the following equation,
\begin{equation}
T_{\rm ex} \ {\rm (K)} \ = \ \frac{5.53}{\ln\{ 1 + 5.53 / (T(^{12}{\rm CO}) + 0.819) \} }.\label{eq1}
\end{equation}
Despite of the strong self-absorption in the $^{12}$CO profiles, observed brightness temperature in the $^{13}$CO clumps is typically very high up to 45\,K in clump III, which corresponds to $T_{\rm ex}$ of 49\,K in equation\,(\ref{eq1}).
This figure is roughly consistent with the LVG analysis, resulting in $T_{\rm k} > 30$\,--\,50\,K (Figure\,\ref{lvg1}) in regions A\,--\,C which are located in the outskirts of the ring cloud.
We therefore adopt a uniform $T_{\rm ex}$ of 49\,K and estimate $M_{\rm LTE}$ of the dense ring structure of the ring cloud in Figure\,\ref{lb+vb}(d) to be $5.2\times10^3$$\msun$ at $^{13}$CO $J$=1--0 integrated intensity $W$($^{13}$CO\,1--0)\,$\geq$\,22.5\,K$\kms$. 
$M_{\rm LTE}$ of the four $^{13}$CO clumps in Figure\,\ref{mopra13+18}(a) are estimated to be $1.0\times10^3$$\msun$ for clump I, $8.8\times10^2$$\msun$ for clump II, $5.7\times10^2$$\msun$ for clump III, and $5.6\times10^2$$\msun$ for clump IV at $W$($^{13}$CO\,1--0)\,$\geq$\,32.5\,K$\kms$ (see also Table\,\ref{table_clump}).
The H$_2$ column density $N$(H$_2$) of the $^{13}$CO clumps is typically as high as $1\times10^{23}$\,cm$^{-2}$.

\subsection{Comparisons with infrared images}
\begin{figure}
\epsscale{.7}
\plotone{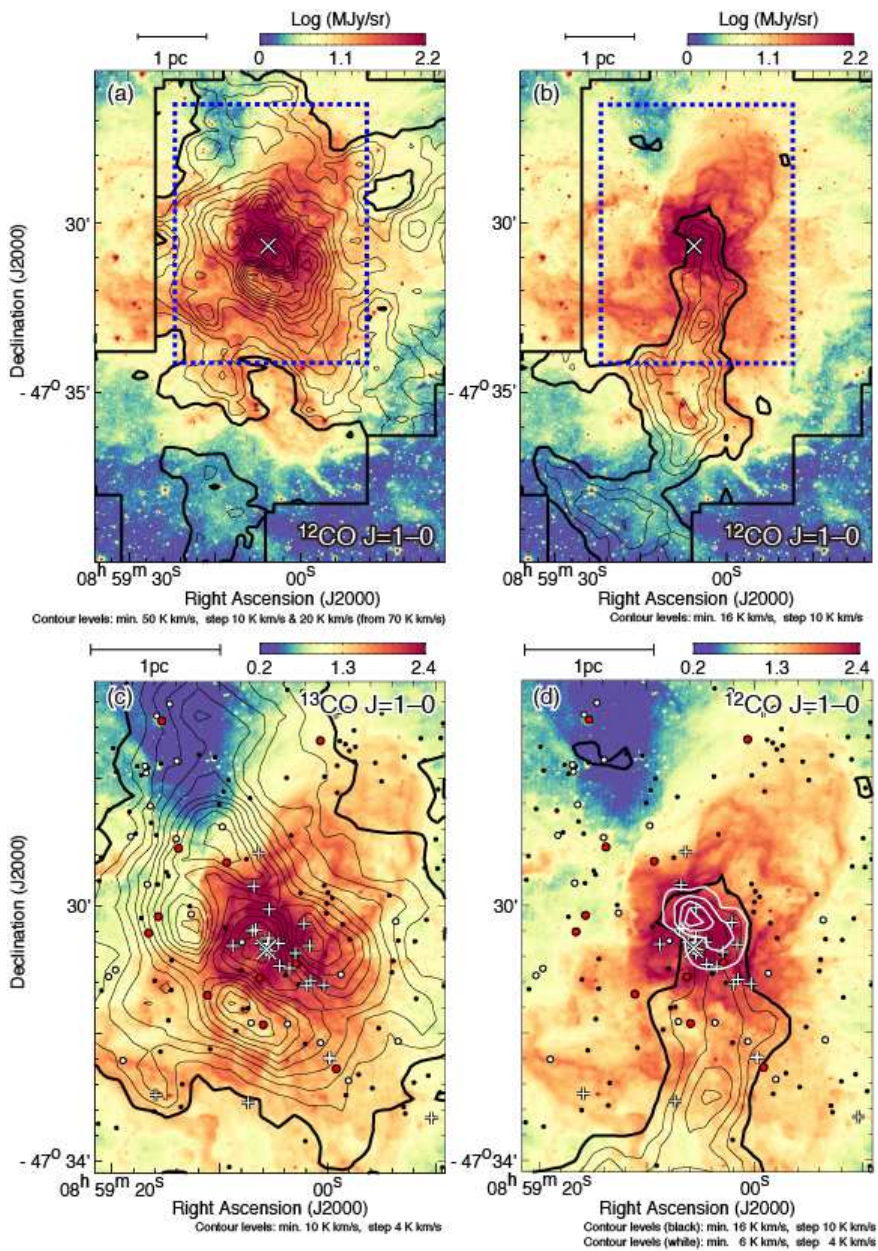}
\caption{Comparisons between the two molecular clouds (contours) and the Spitzer 3.6\,$\mu$m image. The velocity ranges for the left two panels (a) and (c) and for the right two panels (b) and (d) are $-8.5$\,--$+9.1$$\kms$ and $+9.1$\,--$+17.5$$\kms$, respectively. White contours in (d) indicate the bridge feature BR1 integrated over a velocity range $+9.1$\,--\,$11.5$$\kms$. YSOs and O star candidates are plotted with the same manner as in Figure\,\ref{rcw38}.
 \label{spi0}}
\end{figure}

Figure\,\ref{spi0} shows a comparison between the $^{12}$CO $J$=1--0 distribution and the {\it Spitzer} 3.6\,$\mu$m image which consists of several filamentary features around the central peak. Only panel (c) is for $^{13}$CO. 
Figure\,\ref{spi0}(a), a large-scale view, shows that the ring cloud and the intense 3.6\,$\mu$m emission corresponds well with each other. The finger cloud in Figre\,\ref{spi0}(b) shows possible correspondence with the southern part of the 3.6\,$\mu$m distribution at (R.A., Dec.)=($08^{\rm h}\,59^{\rm m}\,5^{\rm s}$, $-47\degr\,35\arcmin$).  
Figure\,\ref{spi0}(c) compares the YSOs and O star candidates in the ring cloud. The YSO distribution corresponds to the overall distribution of the ring cloud, and the O stars with the northwestern part of the cavity of the ring cloud. 
Figure\,\ref{spi0}(d) indicates that nearly 20 of the O stars are located toward the tip of the finger cloud overlapping with the bridging feature, while the distribution of the O stars shows an offset slightly toward the south of the peak of the finger cloud. 
The other YSOs do not show strong correlation with the finger cloud in the south (Figure\,\ref{spi0}(c)).

\begin{figure}
\epsscale{1.}
\plotone{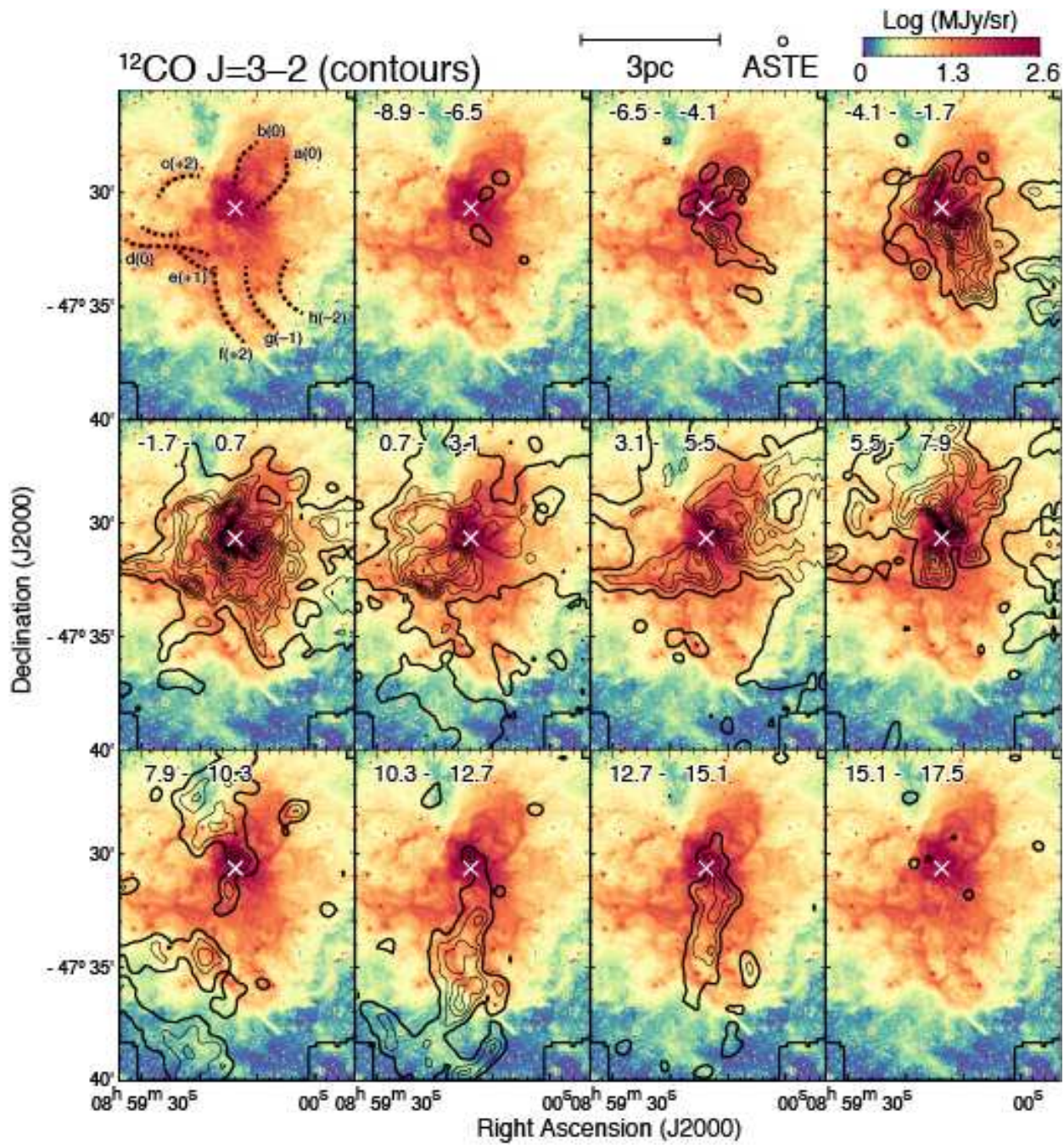}
\caption{Velocity channel maps of the $^{12}$CO $J$=3--2 emission (contours) superimposed on the Spitzer 3.6\,$\mu$m image. In the left-top panel, approximate distributions of the infrared filaments (filaments a\,--\,h) are depicted by the dashed lines. The parenthetic numbers indicate peak velocities of the filaments in km\,s$^{-1}$. (see Table\,2)
 \label{channel}}
\end{figure}

\begin{figure}
\epsscale{.7}
\plotone{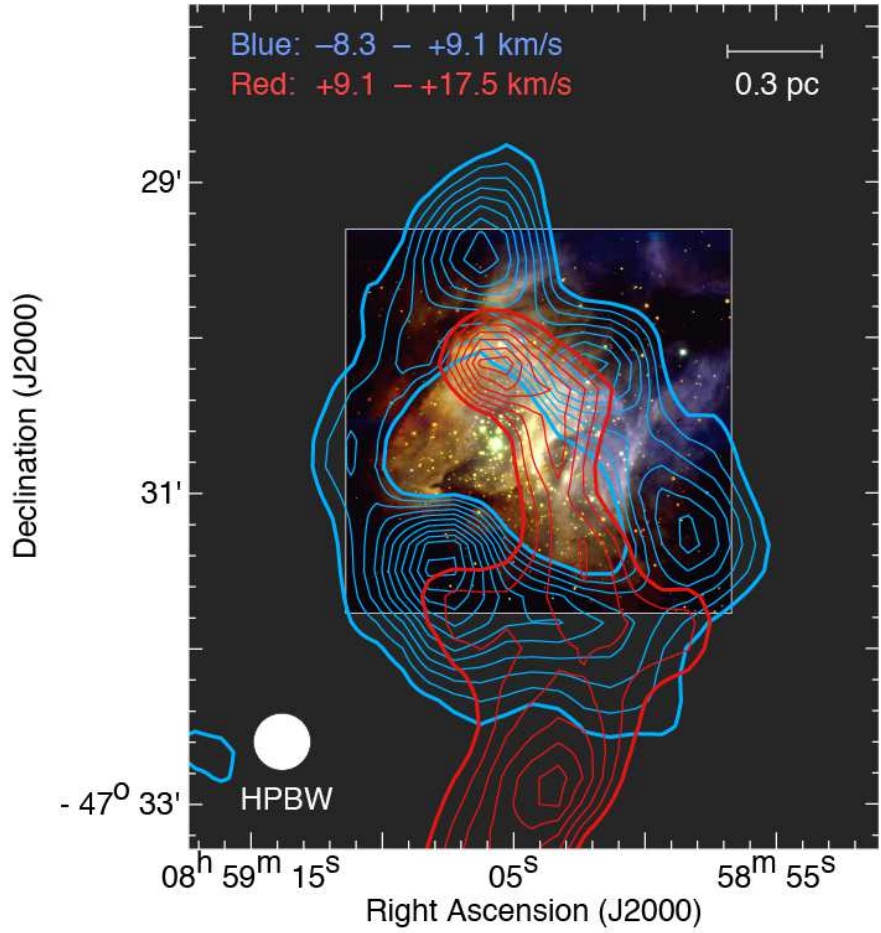}
\caption{$^{12}$CO $J$=3--2 distributions of the ring cloud (blue contours) and the finger cloud (red contours) superimposed on the VLT image in Figure\,\ref{rcw38}. The contours are plotted at every 15\,K\,km\,s$^{-1}$ from 130\,K\,km\,s$^{-1}$ for the ring cloud and at every 5\,K\,km\,s$^{-1}$ from 20\,K\,km\,s$^{-1}$ for the finger cloud.
 \label{vlt+red}}
\end{figure}

Figure\,\ref{channel} shows velocity channel distributions of CO overlaid on the infrared image. We label the nine infrared filamentary features by a--h in the first panel of Figure\,\ref{channel}. The CO features generally show good correspondence with the infrared  features as listed in Table\,2.
Correspondence of the infrared image with the finger cloud is not so obvious in Figure\,\ref{channel}. Figure\,\ref{vlt+red}, an overlay of the ring cloud and the finger cloud with the VLT image at a smaller scale, however, shows a good agreement between the finger cloud and IRS\,1 which is elongated in the north to the south while the angular resolution of the CO distribution is too coarse for spatially resolving the bridging feature to a resolution comparable to the infrared image. We interpret that the dark lane toward IRS\,1 along with another dust feature comprising the western cavity
corresponds to the finger cloud which probably lies on the near side of the cluster. IRS\,1 is clearly irradiated by the cluster and the physical association of the bridging feature with the cluster is supported.

\section{Discussion}

\subsection{RCW\,38 as the youngest super star cluster in the Milky Way}
The number of known super star clusters in the Milky Way is 13, including RCW\,38 and the 12 clusters listed in Table\,2 of \citet{por2010}.
The cluster members of RCW\,38 were previously estimated to be a few 1000 (Wolk et al. 2006) and RCW38 was not included in the table by \citet{por2010} which limited the cluster mass to be more than $10^4$\,$M_\odot$. 
The recent estimate shows that RCW\,38 harbors 10,000 stars \citep{kuh2015} and it is appropriate to include RCW\,38 as one of the super star clusters.
Among the 13 RCW\,38 is the youngest one with an estimated age of less than 1\,Myr \citep[][Sections 4.2 and 4.3 of the present work]{wol2006} and the ages of the others are $\gtrsim2$\,Myrs. The Orion Nebula Cluster is listed as an OB association with an age of 1\,Myr in \citet{por2010}. The ambient interstellar medium will reflect youth of a cluster, and cloud dispersal by the feedback will be less intensive in younger clusters than in more evolved clusters.
According to eye inspection of the infrared images, only five clusters among the 13 clusters, RCW\,38, NGC\,3603, Westerlund\,2, DBS[2003]179, and Trumpler\,14, are associated with dust nebulosity, and their ages are from 2\,Myrs to 3.5\,Myrs except for RCW\,38. 
The remaining 8 clusters, most of which have ages between 3.5\,Myrs and 18\,Myrs, have no associated nebulosity, with an exception of the Arches cluster having an age of 2\,Myrs. It is likely that the interstellar medium around the 8 clusters is completely dispersed by the stellar feedback including ionization and stellar winds.

\begin{figure}
\epsscale{.6}
\plotone{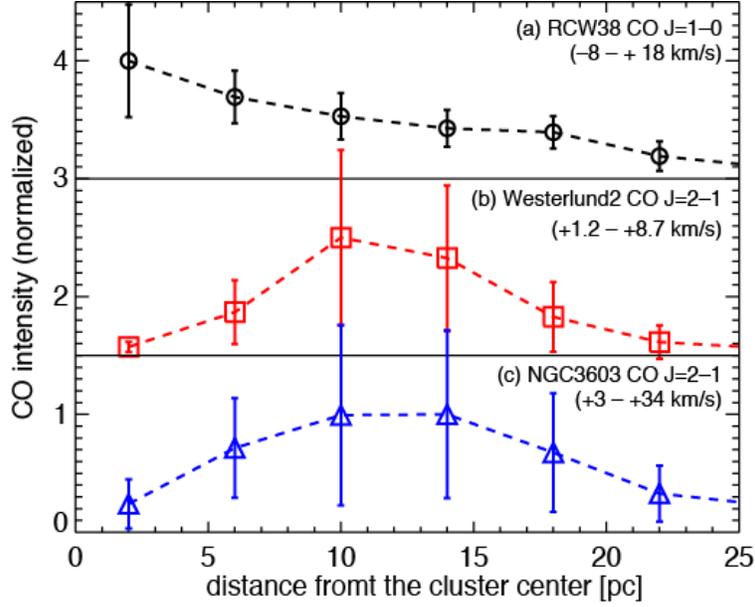}
\caption{Radial distributions of the CO emission in the three SSCs RCW\,38 (a), Westerlund\,2 (b), and NGC\,3603 (c). 
(a) The NANTEN2 CO $J$=1--0 data set shown in Figure\,\ref{nasco_lb} is used. The present velocity range includes the two colliding clouds.
(b) The NANTEN2 CO $J$=2--1 dataset in \citet{fur2009} is used. Here the present velocity range includes only the blue-shifted cloud, because the red-shifted cloud which overlaps the SSC along the line-of-sight is indeed located relatively distant from the cluster compared with the blue-shifted cloud \citep{oha2010}.
(c) The NANTEN2 CO $J$=2--1 dataset in \citet{fuk2014} is used. The velocity range includes the two colliding clouds.
 \label{radial_plot}}
\end{figure}

Figure \ref{radial_plot} shows the averaged CO intensity in the three clusters with nebulosity RCW\,38, NGC\,3603, and Westerlund\,2, and indicates that RCW\,38 still has significant molecular mass within $\sim$3\,pc of the cluster.
In Westerlund\,2 and NGC\,3603 whose ages are 2\,Myrs the central part of the molecular clouds within 10\,pc of the cluster is already strongly ionized/dispersed.
The rich remaining gas toward the cluster is a unique feature of RCW\,38 and is consistent with its age significantly smaller than 2\,Myrs. RCW\,38 may therefore provide a best opportunity to study the initial conditions for the super star cluster formation.

\subsection{The two clouds; physical association and evidence for collision}
The present observations have revealed the distribution of the two molecular clouds with 12$\kms$ velocity separation toward RCW\,38 at a spatial resolution of $\sim$0.1\,pc.
The following pieces of evidence offer robust verification of the association between the two clouds and the cluster. 
\begin{enumerate}
\item The ring cloud shows a central cavity created by the feedback of the cluster.
The ring cloud is also associated with the infrared filamentary/bubble-like features heated by the cluster.
The finger cloud shows good correspondence with IRS\,1 and the western cavity.
The morphological correspondence supports the physical association of the clouds with the cluster.
\item The line intensity ratio of the $^{12}$CO $J$=3--2 emission to $^{12}$CO $J$=1--0 emission shows significantly high values, indicating high gas temperatures toward the cluster. LVG calculations show that the temperature there is higher than 20\,--\,30\,K, consistent with local heating by the O stars in the cluster.
\item Another sign of the connection between the two clouds is seen as the bridging features in velocity between them at least in three places including the direction of the cluster center. Alternative possibility that the bridging features are due to stellar winds is unlikely as discussed later in this section.
\end{enumerate}

In RCW\,38 the observed cloud velocity separation is large 12$\kms$ as seen in the previous two cases Westerlund\,2 and NGC\,3603 \citep{fur2009, fuk2014}. 
The velocity is too large for the clouds to be gravitationally bound by the total mass of the clouds and cluster, less than $10^{5}$\,$M_\odot$. This indicates that the association of the two clouds is by chance. 
We frame a hypothesis that the two clouds collided with each other recently and that the collision triggered formation of stars, primarily O stars, as have been suggested for Westerlund\,2 and NGC\,3603.  
We note that the fourth and fifth super star clusters with nebulosity, DBS[2003]\,179 and Trumpler\,14, are also associated with two molecular clouds which appear colliding to form the cluster (Kuwahara et al. in preparation; Fukui et al. in preparation). 
Therefore, it is possible that the five super star clusters with nebulosity are all formed by triggering under cloud-cloud collision, suggesting the important role of collision in formation of a super star cluster and multiple O stars. In the remaining 8 super star clusters without associated nebulosity it is not likely that we are able to detect the parent molecular clouds.

Stellar wind acceleration may be an alternative interpretation to cloud-cloud collision (e.g., see discussion in \citealt{fur2009}). 
We see a velocity gradient in the finger cloud of RCW\,38 (Section\,3), which may suggest some acceleration by the O stars via the stellar winds. This is however not likely because the gradient is extended outside the cluster and rather uniform over the finger cloud; the velocity shift should be most prominent toward the cluster if the winds originating from the O stars play a role. 
We also note that the finger cloud is located on the nearside of the cluster as suggested by the coincidence with IRS\,1 (Figure\,\ref{vlt+red}), implying that the wind acceleration will cause blue-shift instead of the observed red-shift. The collision between the two clouds is therefore a plausible interpretation, and we exclude wind acceleration as the cause of the two velocities of the clouds.

The possible role of the winds in acceleration raises an alternative to the origin of the bridging feature toward the cluster center. Since the bridging feature, having a tight link with the finger cloud, shows a good correspondence with the O stars (Figure \ref{spi0}d), the bridging feature may be part of the finger cloud accelerated by the stellar winds and not created by the collision. The mass of the bridging feature in Figure\,11d is estimated to be 17$\msun$ from the $^{12}$CO intensity with the X(CO) factor. 
The kinetic energy in the bridging feature is estimated to be $10^{46}$\,erg, small enough to be accelerated by the winds of $\sim$20 O stars whose kinetic energy can be more than $10^{51}$\,erg (e.g., in \citealt{fur2009}). In this scenario, the wind-accelerated gas will be blue-shifted since the finger cloud lies in front of the O stars. This contradicts red-shift of the bridging feature. The velocity range of the bridging feature which lies only between the two velocities of the colliding clouds, not beyond the velocity of the finger cloud, is also consistent with the collision \citep{haw2015b,haw2015,tak2014,tor2015}. The other two bridging features in the south (Figure \ref{comp}) are not associated with O stars and it is unlikely that the winds are a source of acceleration. 
To summarize, the three bridging features are likely formed by the collisional interaction between the ring cloud and the finger cloud, lending additional support for the cloud-cloud collision.

Recent theoretical studies of global numerical simulations of galactic disks indicate that collisions between giant molecular clouds are frequent with a mean free interval of $\sim$10\,Myrs \citep{tas2009,dob2015}. This is comparable to the typical lifetime of a giant molecular cloud 20\,Myrs \citep{fuk1999,yam2001,kaw2009,fuk2010}, suggesting that a giant molecular cloud experiences cloud-cloud collision more than once in its lifetime. It is also possible that the relatively rich molecular gas in the Vela molecular ridge where RCW\,38 is located \citep{yam1999} favors more frequent encounters between the clouds than the average over the Wilky Way.

\subsection{Details of the collisional interaction}
Currently, we see the two clouds, the bridging feature, and the O star cluster, all of which are within 1\,pc of the cluster center. The ring cloud has a cavity of 0.3\,pc radius which is likely ionized by IRS\,2 and the other O stars. The finger cloud is probably being ionized and observed as IRS\,1 whose surface is irradiated by the O stars. Theoretical studies of the collision scenario show that a shock-compressed interface layer is formed between the two clouds \citep{hab1992,ana2010}. It is likely that major part of the interface layer has already been converted to the O stars and that the bridging feature is a remnant of the interface layer.

By assuming tentatively that the relative cloud motion is inclined to the line of sight by 45$^\circ$, we estimate the relative velocity of the two clouds to be 17$\kms$. The actual velocity of the shocked interface layer may be decelerated by momentum conversation in collision between the two clouds \citep[e.g.][]{haw2015b}, and we adopt 10 $\kms$ as the relative velocity of the interface layer.
The distribution of the triggering-formed O stars is extended by about 1\,pc (Figure\,\ref{spi0}).
We estimate the typical collision timescale to be $\sim1\,\textrm{pc} / 10\,\textrm{km}\,\textrm{s}^{-1}\,\simeq\,0.1\,\textrm{Myrs}$, which gives roughly the age of the O stars in the cluster. 

If we assume the O stars are formed by a constant mass accretion rate in the collision time scale $10^{5}$ yrs, the mass accretion rate is estimated to be $4\times10^{-4}\msun$/yr for a 40\,$\msun$ star (an O5.5 star).
Magnetohydrodynamical numerical simulations of cloud-cloud collision have shown that the turbulence is enhanced and the magnetic field is amplified in the interface layer \citep{ino2013}. This leads to two orders of magnitude increase of the mass accretion rate from the pre-collision state; e.g., a core of 126\,$\msun$ is formed in 0.3\,Myrs in the shocked interface layer as shown by model\,4 of Inoue and Fukui (2013) and such a massive core is a promising precursor of an O star. The total mass of the O stars in RCW\,38 is estimated to be $\sim20\times 20=400\msun$, if we assume a typical mass of an O star to be 20$\msun$.
This corresponds to $\sim$10\,\% of the total $^{13}$CO clump mass of the ring cloud 3000$\msun$ (see Table\,1).

The cavity of the ring cloud is created by ionization of the O stars. The propagating velocity of the ionization front is roughly estimated to be 3 $\kms$ from a ratio 0.3\,pc/$10^{5}$yrs.
 This is consistent with the ionization front velocity in a molecular cloud driven by H{\sc ii} regions \citep{spi1968}. In Figure\,\ref{radial_plot} RCW\,38 shows a molecular peak within 3\,pc of the center, whereas the other two NGC\,3603 and Westerlund\,2 show decrease of molecular gas within 10\,pc of the cluster. This trend is explained as due to ionization. The latter two clusters have an age of 2\,Myrs, an order of magnitude larger than RCW\,38. By taking a ratio of 10\,pc/2\,Myrs, we roughly estimate the average velocity of the ionization front to be 5$\kms$, which is comparable to 3$\kms$ in RCW\,38. If we assume velocity of the ionization front to be 3\,--\,5$\kms$ and common molecular cloud density, the ring cloud of 1\,pc radius will be ionized within 0.2\,--\,0.3\,Myrs and triggered star formation will be completely terminated. This poses an upper limit of a few times 0.1\,Myrs for the age spread of the stars formed by triggering. We note that the age spreads of the youngest members of the two super star clusters, NGC\,3603 and Westerlund\,1, are determined to be 0.1\,--\,0.3\,Myrs, an order of magnitude larger than their cluster age 2\,--\,5\,Myrs, by a careful photometric study with VLT for proper-motion selected young cluster members with HST \citep{kud2012}. 
We note that model calculations of NGC\,3603 by \citet{ban2013} and \cite{ban2014} also suggest the rapid cluster formation through a single starburst event followed by significant residual gas explusion.
The small age spread is possibly explained by the collision scenario so that formation of $\sim$20 O stars by triggering is terminated quickly due to the ionization in the order of 0.1\,Myrs.
Most of the low-mass stars were probably formed in the ring cloud prior to the cloud-cloud collision, since they are extended over the ring cloud with a radius of a few pc (Figure\,\ref{rcw38}). We suggest that this naturally causes duality in the stellar age in RCW\,38; i.e., young O stars of 0.1-Myr age and older low-mass stars of 1-Myr age, which is to be observationally confirmed for the cluster members at infrared and X rays. 
Such duality in age is also found in the other super star clusters (e.g., for NGC\,3603 see \citealt{har2008}).

\subsection{Conditions for the multiple O star formation}
It is shown that cloud-cloud collision triggered formation of the O stars in center of the super star cluster RCW\,38. 
The present two clouds show a sign of collisional interaction, the bridging features, in the three regions denoted as BR1, BR2 and BR3 in Figure\,5.
Only BR1 shows formation of multiple O stars with projected density of $\sim$30\,(pc)$^{-2}$ $[=15/(1.0 \mathrm{pc}\times0.5 \mathrm{pc})]$, and BR2 and BR3 no O star formation. The ring cloud has the highest molecular column density $1\times 10^{23}$\,cm$^{-2}$ around BR1 and has lower column density of $(1-2)\times 10^{22}$\,cm$^{-2}$ in BR2 and BR3. On the other hand, the finger cloud has nearly uniform column density of $1\times 10^{22}$\,cm$^{-2}$, and the relative velocities between the two clouds are almost the same in the three regions.
We suggest that the high column density is the necessary condition for O star formation in cloud-cloud collision. The total molecular masses toward the three regions are estimated to be $\sim$260$\msun$, $\sim$90$\msun$, and $\sim$100$\msun$, respectively, suggesting that BR2 and BR3 are probably not massive enough to form multiple O stars like in BR1, while it is possible that the collision is in its early phase prior to O star formation in BR2 and BR3.

Table\,3 compares the observational properties of the molecular clouds in RCW\,38 with the other two super star clusters, Westerlund\,2 \citep{fur2009,oha2010} and NGC\,3603 \citep{fuk2014}, and with the single O stars in M\,20 \citep{tor2011}, and RCW\,120 \citep{tor2015}, where cloud-cloud collision is suggested to be a trigger of O star formation. The comparison shows a trend that the multiple O star formation as in RCW\,38 is found only for a high column density of $10^{23}$\,cm$^{-2}$ in either of the two clouds, whereas the cloud mass and relative velocity do not show a particular trend in multiple O star formation; i.e., the mass range is from $10^{3}\msun$ to $10^{5}\msun$ and the observed relative velocity is 10\,--\,30$\kms$ in either of the multiple or single O star formation. The high column density in one of the two colliding clouds is therefore possibly the necessary condition for the multiple O star formation.

The two scenarios for high-mass star formation, the monolithic collapse and competitive accretion, are not yet verified by confrontation with observations \citep{tan2014}. Cloud-cloud collision, another possible scenario for high-mass star formation, provides directly testable three observational signatures including \underline{(1) two clouds associated with young O star(s)} with \underline{(2) relative velocity greater than 10 $\kms$,} and \underline{(3) the bridging feature} between the two clouds, and RCW\,38 show all the signatures. 

According to \citet{ino2013}, the effective sound speed in the shocked interface layer is isotropic and given roughly as the velocity separation of the two colliding gas. 
It is then possible that low velocity collision of molecular gas may lead to a lower mass accretion rate and formation of a lower mass star. 
Observations of low-mass star formation without O stars triggered by cloud-cloud collision are presented by several authors \citep{dua2010,hig2010,nak2012}. 
The relative velocity of these collisions is as small as a few km\,s$^{-1}$, and leaves room for non-triggered star formation under a gravitationally bound system.It remains to be explored what is the role of cloud-cloud collision in low-mass star formation.

\section{Conclusions}
We have carried out CO $J$=1--0 and $J$=3--2 observations toward the super star cluster RCW\,38 with Mopra, ASTE and NANTEN2 mm/sub-mm telescopes. The main conclusions of the present study are summarized as follows;
\begin{enumerate}

\item We have revealed distributions of two molecular clouds at velocities of 2$\kms$ and 14$\kms$  toward RCW\,38.
The ring cloud (the 2$\kms$ cloud) shows a ring-like shape with a cavity ionized by the cluster and has a high molecular column density of $\sim10^{23}$\,cm$^{-2}$.
The other, the finger cloud (the 14$\kms$ cloud), has a tip toward the cluster and has a lower molecular column density of $\sim10^{22}$\,cm$^{-2}$. 
The total masses of the ring cloud and the finger cloud are $3.0\times10^4\msun$ and $1.9\times10^3\msun$, respectively.

\item It is likely that the two clouds are physically associated with the cluster as verified by the high intensity ratio of the $J$=3--2 transition to the $J$=1--0 transition, $R_{3-2/1-0}$, toward the cluster.
The observed ratio indicates kinetic temperature of $\gtrsim20$\,--\,30\,K according to the LVG calculations, which is significantly higher than the canonical molecular cloud temperature 10\,K with no extra heat source. 
The heating is mainly due to the O stars.
The association is further supported by the distribution of the molecular clouds corresponding to the cluster and the infrared dust features; the ring cloud shows the cavity toward the cluster center and good coincidence with the extended dust features, and the finger cloud corresponds to the infrared ridge IRS\,1.
In addition, the two clouds are linked with each other by bridging features in velocity at least in three places including the direction of the cluster, supporting further the physical connection between the two clouds.

\item The total mass of the clouds and the cluster is less than $10^5$\,$M_\odot$. The velocity separation of 12$\kms$ is too large for the clouds to be gravitationally bound, and we suggest that the clouds encountered by chance.
We present an interpretation that the two clouds collided 0.1\,Myrs ago at 17\, km\,s$^{-1}$ with each other to trigger formation of the O stars in the cluster. 
We argue that the double O5.5 star (IRS\,2) was formed within this timescale at an average mass accretion rate $\sim4\times10^{-4}$\,$M_\odot$\,yr$^{-1}$.

\item The tip of the finger cloud and the bridging feature connecting the two clouds well coincide with the O stars within 0.5\,pc of the cluster center, indicating that the triggering happened only toward the inner 0.5\,pc and the other member low-mass stars outside 0.5\,pc are pre-existent, being extended over the ring cloud where collisional interaction is not taking place.
RCW\,38 is the third super star cluster alongside of Westerlund\,2 and NGC\,3603 where cloud-cloud collision triggered O star formation in a super star cluster, lending further support for an important role of supersonic collision in O star formation. 
Among the three super star clusters, RCW\,38 is unique because it is the youngest cluster where the initial conditions prior to the O-star formation still hold without significant cloud dispersal by stellar feedbacks beyond 1\,pc.

\item The present findings suggest a possibility to prepare a recipe for O star formation in a super star cluster. 
Multiple O star formation is triggered where collision of two clouds at velocity 10\,--\,30$\kms$ takes place. 
One of the clouds has a high column density of $\sim10^{23}$\,cm$^{-2}$, while the other has a lower column density. 
Since O star formation takes place only at the spots of the collision, the distribution of O stars keeps the distribution of the collisional interaction as long as the gravitational rearrangement of O stars after the collision does not play a role. 
This picture provides a natural set up for the highly turbulent interstellar medium as the initial condition for O  star formation along the line of suggestion by \citet{zin2007}.

\end{enumerate}

In order to better establish the role of cloud-cloud collision in O star formation, we need an extensive systematic study of molecular clouds toward a number of young H{\sc ii} regions and young stars whose mass exceeds 20\,$M_\odot$. 
Such a study is on-going and will allow us to test how cloud-cloud collision triggers and regulates O star formation. 
It is imperative to attain very high angular resolution in imaging the molecular gas to resolve dense cores in the shocked region and ALMA is the most important instrument toward the goal.
 It is also a challenge to test if mini-starbursts in galaxies are triggered by supersonic collisions between molecular clouds.

\acknowledgments

The authors thank the anonymous referee for his/her helpful comments. This work was financially supported by Grants-in-Aid for Scientific Research (KAKENHI) of the Japanese society for the Promotion of Science (JSPS; grant numbers 15H05694, 15K17607, 24224005, 26247026, 25287035, and 23540277). NANTEN2 is an international collaboration of ten universities, Nagoya University, Osaka Prefecture University, University of Cologne, University of Bonn, Seoul National University, University of Chile, University of New South Wales, Macquarie University, University of Sydney and Zurich Technical University. The ASTE telescope is operated by National Astronomical Observatory of Japan (NAOJ).


\clearpage

\clearpage

\begin{table}
\begin{center}
\caption{Parameters of the $^{13}$CO clumps I\,--\,IV}
\label{table_clump}
\begin{tabular}{cccccccc}
\tableline\tableline
 \multirow{2}{*}{Name}	&	R.A. 	& Dec. &  $v_{\rm LSR}$	& $\Delta v$	& size  &  $M_{\rm LTE}$ \\
 & [$^{\rm h\, m\, s}$]	& [$^\circ$\,$'$\,$''$]	& [km\,s$^{-1}$]	& [km\,s$^{-1}$] 	&[pc]		& [M$_\odot$] \\
 (1)&(2)&(3)&(4)&(5)&(6)&(7)\\
\tableline

clump I		& 8:59:9.2		& $-47$:29:1.5		& 5.0		& 4.2		& 0.5 	& $1.0\times10^3$\\
clump II		& 8:59:12.5	& $-47$:30:16.9	& 1.7		& 4.1		& 0.4		& $8.8\times10^2$ \\
clump III		& 8:59:8.4		& $-47$:31:29.7	& 2.3		& 5.3	 	& 0.3		& $5.7\times10^2$ \\
clump IV		& 8:59:1.2		& $-47$:30:27.3	& 2.3		&4.8		& 0.4		& $5.6\times10^2$	 \\

\tableline
\end{tabular}
\tablecomments{Column (1): Name. (2, 3): Position in J2000. (4, 5): Peak $v_{\rm LSR}$ and velocity width $\Delta v$ (FWHM) of the $^{13}$CO $J$=1--0 profile, where fitting with a Gaussian function is used to derive the parameters. (6): Size of the $^{13}$CO clump, measured as geometric average of major and minor diameters at the $^{13}$CO integrated intensity 45\,K\,km\,s$^{-1}$. (7) Molecular mass within the clump size.}
\end{center}
\end{table}

\clearpage

\clearpage

\clearpage

\begin{table}
\begin{center}
\caption{CO velocity range corresponding to the infrared filaments in Figure\,\ref{channel}}
\label{tab3}
\begin{tabular}{cc}
\tableline\tableline
 \multirow{2}{*}{Name}	&	$v_{\rm LSR}$ range	 \\
	&  [km\,s$^{-1}$] \\
 (1)	&  (2)\\
 \tableline
 a	& $-1.7$\,--\,+0.7 \\
 b	& $-1.7$\,--\,+0.7 \\
 c	& $+0.7$\,--\,+3.1 \\
 d (two filaments)	& $-1.7$\,--\,+0.7 \\
 e	& $+0.7$\,--\,+3.1 \\
 f	& $-0.7$\,--\,+3.1 \\
 g	& $-1.7$\,--\,+0.7 \\
 h	& $-4.1$\,--\,-1.7 \\
\tableline
\tableline
\end{tabular}
\tablecomments{Column: (1) Name of the filament. (2) The velocity range in which CO emission corresponds to the filament.}
\end{center}
\end{table}

\begin{table}
\footnotesize
\begin{center}
\caption{Comparison of five regions of cloud-cloud collision.}
\label{tab3}
\begin{tabular}{cccccc}
\tableline\tableline
 \multirow{2}{*}{Name}	&	cloud masses	&   column densities	&   velocity separation	&   \multirow{2}{*}{\# of O stars}	&  \multirow{2}{*}{reference} \\
					&  [$M_\odot$]		&  [cm$^{-2}$]		&   [km\,s$^{-1}$] 	&&\\
 (1)	&  (2)  & (3)  &  (4)  &  (5)  &  (6)  \\
 \tableline
RCW\,38    & ($2\times10^4$, $3\times10^3$)  & ($1\times10^{23}$, $1\times10^{22}$) & 12 & $\sim20$& This study \\
NGC\,3603 & ($7\times10^4$, $1\times10^4$) & ($1\times10^{23}$, $1\times10^{22}$) & 20 & $\sim30$ & [1] \\
Westerlund\,2 & ($8\times10^4$, $9\times10^4$) & ($2\times10^{23}$, $2\times10^{22}$) & 13 & 14 & [2, 3]\\
M\,20 & ($1\times10^3$, $1\times10^3$) & ($1\times10^{22}$, $1\times10^{22}$) & 7 & 1 & [4]\\
RCW\,120 & ($4\times10^3$, $5\times10^4$) & ($8\times10^{21}$, $3\times10^{22}$) & 20 & 1 & [5] \\

\tableline
\tableline
\end{tabular}
\tablecomments{Column: (1) Name. (2, 3) Molecular masses and column densities of the two colliding clouds. (blue-shifted cloud, red-shifted cloud) (4) Relative radial velocity between the two clouds. (5) Number of O stars created via cloud-cloud collision. (6) References: [1] \citet{fuk2014}, [2] \citet{fur2009}, [3] \citet{oha2010}, [4] \citet{tor2011}, [5] \citet{tor2015}. }
\end{center}
\end{table}

\end{document}